\documentclass[usenatbib]{mn2e}
  \usepackage[T1]{fontenc}
  \usepackage[dvips]{graphicx,color}
  \usepackage{subfigure}
  \usepackage{afterpage}
  \usepackage{times}
\newcommand{\kms} {\mbox{\rm km$\;$s$^{-1}$}}

\title[The ghost of Ly-$\alpha$]
{A new sample of broad-absorption-line quasars exhibiting the ghost of lyman alpha}

\author[M. North, C. Knigge \& M. Goad]  
{Matthew North$^1$\thanks{email: mnorth@star.ucl.ac.uk},
Christian Knigge$^2$ and Mike Goad$^3$\\
$^1$Department of Physics and Astronomy, 
University College London, Gower Street, London WC1E~6BT,~UK\\
$^2$ School of Physics and Astronomy, Highfield, University of Southampton, SO17~1BJ, UK\\
$^3$ Department of Physics and Astronomy, University of Leicester, University Road, LE1~7RH, UK}

\date{Dates to be inserted}

\begin{document}

\maketitle

\begin{abstract}
   We have searched the broad-absorption-line quasar (BAL QSO) sample
   presented by Reichard et~al.\ for objects exhibiting the so-called
   `ghost of Lyman alpha'. This ghost manifests as a hump near
   $-$5900~\kms\ in the troughs of the broad absorption lines and
   provides strong evidence for the importance of line-driving in
   powering the outflows from BAL QSOs.  Of the 224 sample BAL QSOs
   selected from the Sloan Digital Sky Survey (SDSS) Early Data
   Release (EDR), 198 satisfy our redshift constraints and 58 show
   clear evidence of multiple-trough (MT) structure in the C$\;${\sc
   iv}~1550\AA\ line. A composite spectrum constructed from this MT
   sample already shows evidence for a ghost feature. Narrowing our
   classification scheme further, we define a set of 36 objects that
   individually show evidence of a ghost feature, and then apply
   further cuts to arrive at a final `best sample' that contains our seven
   strongest ghost candidates. A further five objects show evidence for a
   ghost feature that is almost strong enough to merit inclusion in
   our best sample. Despite its limited size, our best sample more
   than doubles the number of known BAL QSOs with clear ghost
   signatures and should make an excellent basis for detailed
   follow-up studies.

\end{abstract}

\section{Introduction}

Approximately 10--20\%\ of quasi-stellar objects (QSOs) display
strong, broad, blue-shifted absorption lines in their spectra (Foltz
et~al.\ 1990, Weymann et~al.\ 1991, Reichard et~al.\ 2003b, Hewett \&\ Foltz 2003). 
These sources, the so-called
broad absorption-line (BAL) QSOs, are predominantly radio quiet (Stocke et~al.\
1992) with the majority (85\%; Sprayberry \& Foltz 1992) displaying
strong broad absorption troughs only in lines of high ionization
(HiBALs), e.g., N$\;${\sc v} (1240{\AA}), Si$\;${\sc iv} (1400{\AA}),
C$\;${\sc iv} (1550{\AA}). The remainder exhibit additional BALs in
lines of low-ionization species (LoBALs), most notably Mg$\;${\sc ii} (2800
{\AA}) (Weymann et~al.\ 1991). The emission-line properties of BAL QSOs
and non-BAL QSOs appear identical, while their continua differ only in
their power-law indices and degree of reddening, suggesting that they
are drawn the same parent population (Reichard et~al.\ 2003b). A
straightforward interpretation of these differences is that BAL QSOs
are simply broad-emission-line objects viewed at a particular
orientation. Indeed, in the context of unified models, aside from
differences in radio power, orientation is the key to unifying all AGN
classes.

The broad absorption troughs have long been regarded as signs of
large-scale outflows or winds, whose velocities (as inferred from the
widths of the troughs) can reach 0.1--0.2$c$ (Korista et~al.\
1992). These outflows remove mass, energy and momentum (both linear
and angular) from the QSO and deposit them in the host galaxy. As a
result, they can significantly affect the evolution of the QSO and the
chemical enrichment of its host, for example. Consequently, analyses
of the physics, as well as the overall statistics, of BAL QSOs are
also important to studies of QSOs and AGN more generally.  An
observational feature that has provided much insight into the physics
of outflows from BAL QSOs is the so-called `ghost of Ly $\alpha$'
(Arav et~al.\ 1995). This term refers to a hump near $-$5900~\kms\ seen in
the troughs of the broad absorption lines of some BAL QSOs. This local
maximum can be explained naturally if the outflow is radiatively
accelerated via resonance-line scattering (see Arav 1996 and
references therein for details). Briefly, according to this model, the
ghost is produced when Ly-$\alpha$ broad emission line (BEL) photons
are resonantly scattered by N$\;${\sc v} ions in regions of the
outflow that are moving at $-$5900~\kms\ (relative to the Ly $\alpha$
emission line region). These scatterings transfer momentum and thus
accelerate the wind locally, causing a decrease in the optical depth
at $-$5900~\kms\ in velocity space. Thus observers viewing the QSO
through the outflow will see an increase in flux at this velocity
within the BAL troughs. Furthermore, the profile of this feature
directly reflects the profile of the Ly $\alpha$ BEL; hence the
resultant feature is appropriately named `the ghost of Ly $\alpha$'. 

If this picture is correct, the ghost feature is a direct signature of
the wind driving mechanism and can be used to study the physics
governing the outflows from QSOs.  Despite the potential significance
of the ghost of Ly $\alpha$ for our understanding of (BAL) QSOs, the
set of four objects discussed by Arav (1996) is currently still the only
observational sample of BAL QSOs exhibiting clear ghost signatures. It
is the purpose of this paper to expand this sample. We note from the
outset that our selection method is purely observational and thus
differs from that used by Arav (1996), who based his selection on
criteria derived directly from the radiative-driving model. Our goal
here is simply to construct a new empirical sample of strong `ghost
candidates', based only on the appearance of their C$\;${\sc iv} (and,
in some cases, Si$\;${\sc iv}) BALs.  We do check (Section~\ref{Sec06}) that
none of the objects in our final sample violate the criteria set out
by Arav. However, in many cases the wavelength coverage of the SDSS
data we use is insufficient to confirm that all of the criteria are
satisfied. We therefore defer detailed comparison of the sample
properties to the predictions of the line-driving model for future
investigations.

\section{Data}

We require a set of BAL QSOs not previously scrutinized for ghost
signatures. With the recent data releases from the Sloan Digital Sky
Survey (SDSS), we have access to an unprecedented number of QSOs,
including many BAL QSOs. As a first step, here we take as our parent
sample the BAL QSO catalogue presented by Reichard et~al.\ (2003),
which is based on the SDSS Early Data Release (EDR).  For the purpose
of our analysis, we have chosen to select only objects whose spectra
fully cover the C$\;${\sc iv} BEL and its associated BAL. The
C$\;${\sc iv} BAL tends to display a particularly deep, well-defined
trough and is thus the most likely BAL to exhibit a clear ghost
feature. Thus, given the wavelength coverage of the SDSS 1-D spectra,
a suitable redshift window of $1.66<z<4.94$ was identified\footnote{We
adopt the redshifts given by Reichard et al.\ (2003) unless otherwise
noted.}.  Reichard et~al.'s (2003) sample, almost inevitably, covers this
nicely and, out of a possible 224, immediately provided us with 198
suitable BAL QSOs.  The 1-D spectra of these objects were extracted
directly from the SDSS web site, using the on-line data query
form. These spectra are fully reduced, wavelength-calibrated,
sky-subtracted and corrected for galactic extinction.

\section{Methodology}
\label{SecMeth}

The approach taken in this study is to progressively sub-divide our
BAL QSO sample into sensible categories, at each step eliminating
those objects that do not show convincing ghost signatures. For each
sub-sample, we produce a composite spectrum to highlight structure
that is common to objects across the sample.

\begin{figure}
\includegraphics[scale = 0.4]{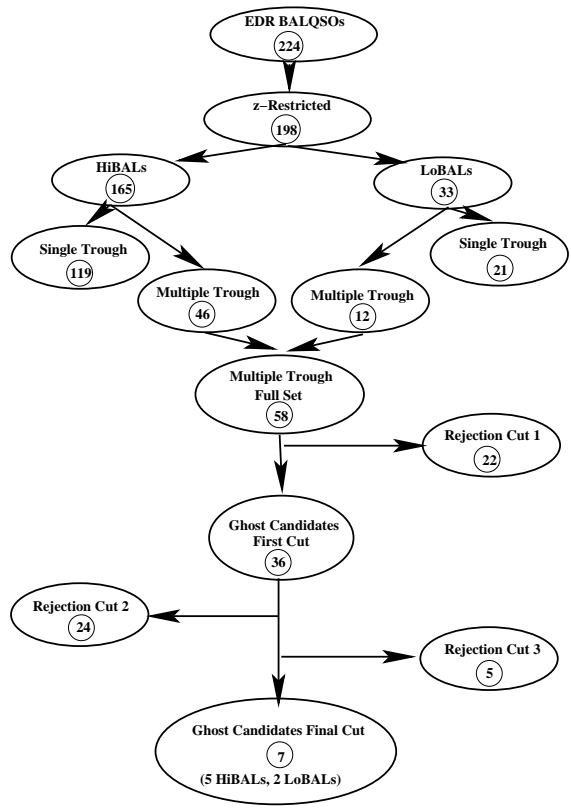}
\caption[Figure 1]{
A simple flow diagram illustrating the method of sub-division employed in this chapter.
}
\label{Fig01}
\end{figure}

Figure~\ref{Fig01} illustrates our attempt to sub-divide the full BAL
QSO sample into manageable and, hopefully, more revealing data
sets. Starting from the complete set of BAL QSOs contained within our
redshift window, we first create two subsets, namely, the
high-ionization BAL QSOs (HiBALs) and low-ionization BAL QSOs
(LoBALs). We adopt Reichard et~al.'s `by eye' classifications for this
purpose with his FeLoBALs being classified as LoBALs. The primary
reason for this step is that it is more common to see highly
structured BALs in LoBALs. One might therefore expect it to be more
difficult to find clear ghost signatures amongst individual LoBALs.

The next sub-division splits these samples into groups exhibiting
single trough (ST) and multiple trough (MT) BALs. Our working
definition of a MT BAL is simply that it should exhibit more than one
clear minimum in its absorption trough. This classification process
was done by eye. The justification for this division is that all ghost 
candidates must exhibit MTs, as by definition the ghost feature is a 
local maximum in a BAL thus dividing what would have been a single
trough into a double trough. As expected, the proportion of MTs is
somewhat higher amongst the LoBALs (36\%) than the HiBALs (28\%).
Even though we expect it to be more difficult to find convincing ghost
candidates amongst the LoBAL MT set, we feel it is nevertheless
important to inspect all BAL QSO exhibiting MTs for this feature. We
therefore re-merge the MT HiBAL and LoBAL sets before making our final
rejection cuts.

In order to help us distinguish likely ghost features from other BAL
structure, we now introduce the concept of a `ghost zone' (GZ). This
is defined as the region in velocity space within which we would
expect to see the peak of the ghost signature. More specifically, we
define the limits of the GZ by the most extreme combinations of the
doublet structures that give rise to the signature itself (see Arav 1996 for further discussion).
Thus we first calculate the rest-frame velocity differences between Ly$\alpha$ 
and each of the two N$\;${\sc v} doublet transitions. We then locate where in the spectra
these velocities occur for each of the doublet pairs in the BELs/BALs of interest
(C$\;${\sc iv} [always] and Si$\;${\sc iv} [where available]). The
maximum and minimum values of these locations fundamentally define the
edges of the GZs for Si$\;${\sc iv} and C$\;${\sc iv}.  We finally
slightly expand the GZs to allow for redshift errors. These are larger
for BAL QSOs than `ordinary' QSOs, in part due to the blue wing absorption of the BEL,
 with typical values (statistical + systematic) around $\Delta{z} \simeq 0.01$ 
(Donald Schneider, personal communication;
see also Schneider et~al.\ 2002). In practice, we actually expect
uncertainties to be more constant in velocity than in redshift.
We therefore expand the GZs by multiplying the limiting wavelengths by a factor $[1 \pm
\Delta{z}/(1+z_{\rm med})] = 1 \pm 0.0032$, where $z_{\rm med} = 2.11$ is the median
redshift of the full BAL QSO sample, and the positive and negative signs refer to
the red and blue edges of the GZ, respectively.

We now pare down our full MT sample by carrying out three rejection
cuts. In the first cut, we remove what we consider obvious non-ghosts
and weak candidates. Thus, in this cut, we reject objects exhibiting
bumps well away from `ghost' velocities; sources with particularly low S/N spectra; and
sources with highly structured BALs. At the end of this iteration, we are
left with our `Rejection Cut 1' (RC1) and `Ghost Candidates
First Cut' (GC1) samples.

\begin{figure*}
\includegraphics[scale = 0.65,angle=270]{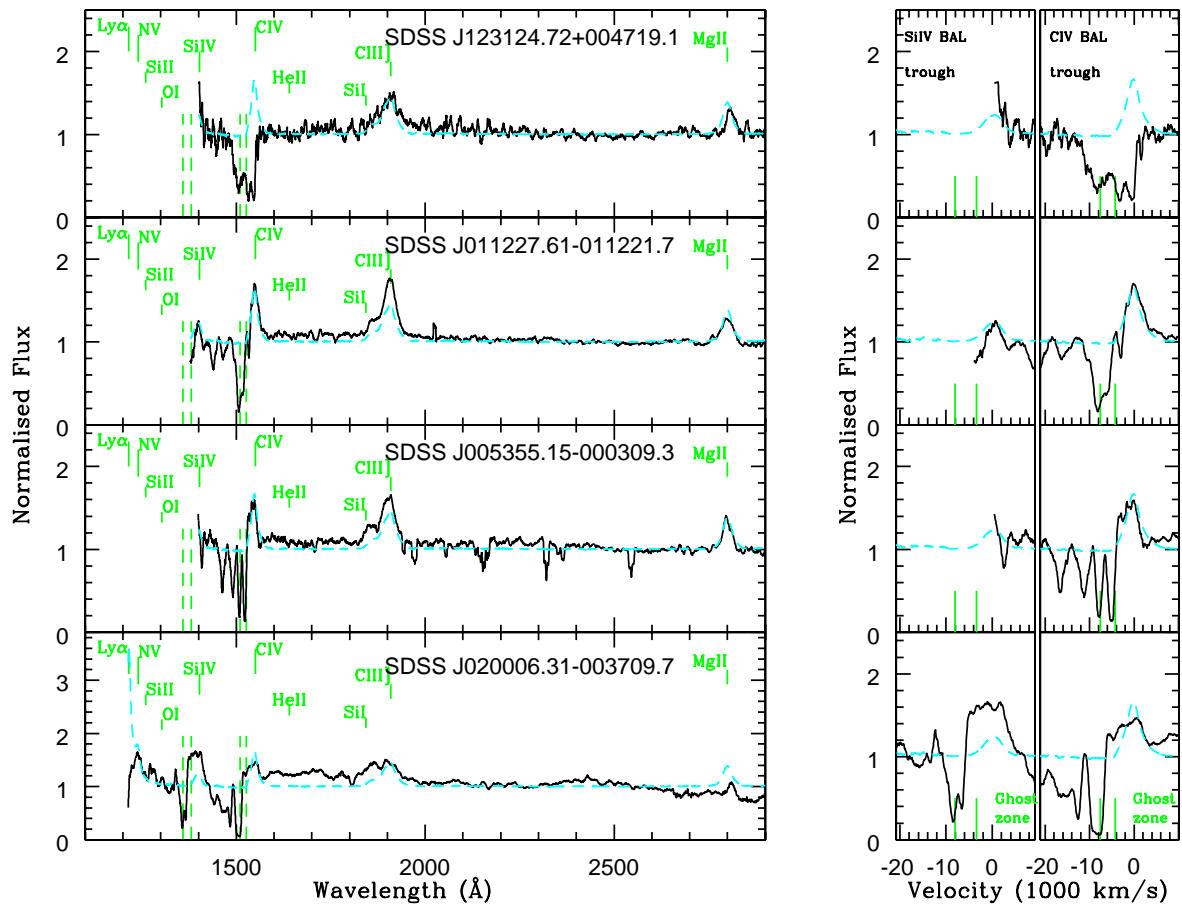}
\caption[Figure 2]{
Examples of MTBAL QSOs rejected at cuts 1 \& 2. The two upper
panels show objects selected from the RC1 sample, whilst the lower two
panels show objects selected from the RC2 sample. The black lines
represent the normalized spectra. The blue dashed lines display the
EDR QSO sample composite. The panels on the right display the
C$\;${\sc iv} and Si$\;${\sc iv} BALs in more detail; see text for further
details.
}
\label{Fig02}
\end{figure*}

The second and third rejection cuts are necessarily more subjective
and designed to leave a final sample that contains only sources with
clear, strong, local maxima in their BALs that appear well within
their GZs. The corresponding sets of rejected objects (`Rejection Cut
2' [RC2] and `Rejection Cut 3' [RC3]) thus contain objects with more
than one local maximum in their BALs, features at/beyond the edges of
the GZ, etc. The distinction between RC2 and RC3 is simply that the
RC3 sample comprises only those objects that narrowly missed inclusion
in our best sample. We thus regard the five objects in RC3 as fairly
strong ghost candidates in their own right. The set of seven objects that
pass all of our rejection cuts comprise our best sample of ghost
candidate (`Ghost Candidates Final Cut' [GCF]).

\begin{figure*}
\includegraphics[scale = 0.65,angle=270]{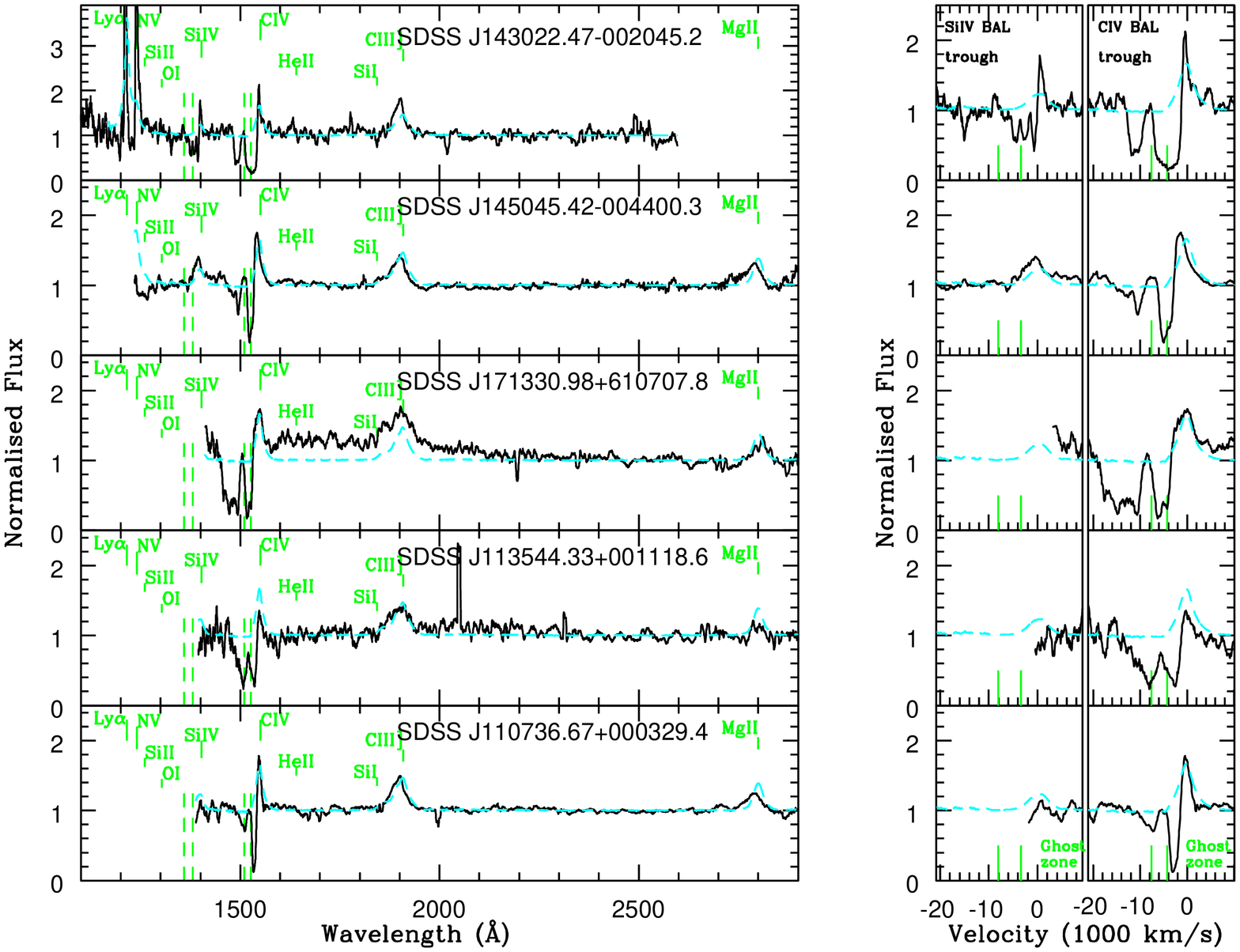}
\caption[Figure 3]{
Plots of the Rejection Cut 3 sample. The black
lines represent the normalized spectra. The blue dashed lines display
the EDR QSO sample composite. The panels on the right display the
C$\;${\sc iv} and Si$\;${\sc iv} BALs in more detail. The vertical
dashed green lines mark out the GZs in both the C$\;${\sc iv} and
Si$\;${\sc iv} BALs.
}
\label{Fig03}
\end{figure*}

Since the three rejection steps outlined above are the most subjective
aspects of our selection process, we show in Fig.~\ref{Fig02} selected spectra
from the RC1 \& RC2 samples, and in Fig.~\ref{Fig03} all spectra for the RC3
sample. These figures illustrate the sort of decisions we were faced
with, and we now briefly describe our reasoning in making these
decisions.

The top two panels in Fig.~\ref{Fig02} show objects selected from the RC1
sample. SDSS~J123124.71+004719.1 displays a feature in the C$\;${\sc
iv} GZ, but is rejected at this stage due to the additional
structure in its BAL and the relatively low S/N ratio of its
spectrum. SDSS~J011227.60$-$011221.7 is clearly an MT BAL QSO, but no
feature is located in or near to the GZ of C$\;${\sc iv}.

The bottom two panels in Fig.~\ref{Fig02} show spectra drawn from the RC2
sample. SDSS~J005355.15$-$000309.3 shows a clear local maximum centred in
the GZ of C$\;${\sc iv}. However, there are also two other local
maxima blueward of this potential ghost feature. In fact, this object
might not be a BAL QSO at all, and the apparent BAL structure might
instead be due to multiple narrow absorbers.  SDSS~J20006.31$-$003709.7
also has a clear feature in the C$\;${\sc iv} BAL, but, for the
redshift adopted by Reichard et~al.\ (2003), it is well outside the
C$\;${\sc iv} GZ. We have nevertheless classified this as RC2 (rather
than RC1), since the flattened appearance of the BELs casts some
doubts over the reliability/accuracy of the redshift estimate for this
object.

Finally, we comment briefly on the RC3 spectra shown in Fig.~\ref{Fig03}. SDSS~J143022.47$-$002045.2, SDSS~J145045.42$-$004400.3 and SDSS~J171330.98+610707.8
all display deep, broad BAL troughs with significant features right on
the blue edges of their respective GZs. The off-centre location of the
features is the reason for their rejection, but clearly all are
nevertheless reasonable ghost candidates. SDSS~J113544.33+001118.6
appears to have a feature that is more closely centred in the
C$\;${\sc iv} GZ. The only reason for its inclusion in RC3 rather than
GCFC is the fact that the spectrum is relatively noisy. Finally, SDSS~J110736.67+000329.4 also has a feature in the C$\;${\sc iv} GZ, but
this feature is of similar size as those found to the red of the
Si$\;${\sc iv} BEL.

Indeed, it is worth emphasizing again at this point that our main goal here is
simply to construct a sample of BAL QSO showing particularly strong
and convincing ghost of Ly $\alpha$ feature in their spectra. By its
nature, this sample is not complete in any statistically meaningful
sense. In particular, we believe the rejection cuts leading to our GCF
sample are quite conservative. Thus our rejection samples may contain
additional ghost candidates, and the RC3 sample, in particular,
contains objects that missed inclusion in our best GCFC sample by only
the narrowest of margins.

\section{Composite spectra}

In this section, we present and discuss composite spectra we have
produced for our various samples. Our rationale for producing these
composites is that they allow us to search for spectral features that
are common to a significant fraction of a given sample. This is
useful, since, if BAL QSO outflows are radiatively driven, ghosts
should be more common than other types of BAL structure. We may then
expect to see ghosts even in composites constructed from samples that
have not been specifically selected for displaying this feature
(i.e., samples high up in the hierarchy in Fig.~\ref{Fig01}).

\begin{figure*}
\includegraphics[scale = 0.65,angle=270]{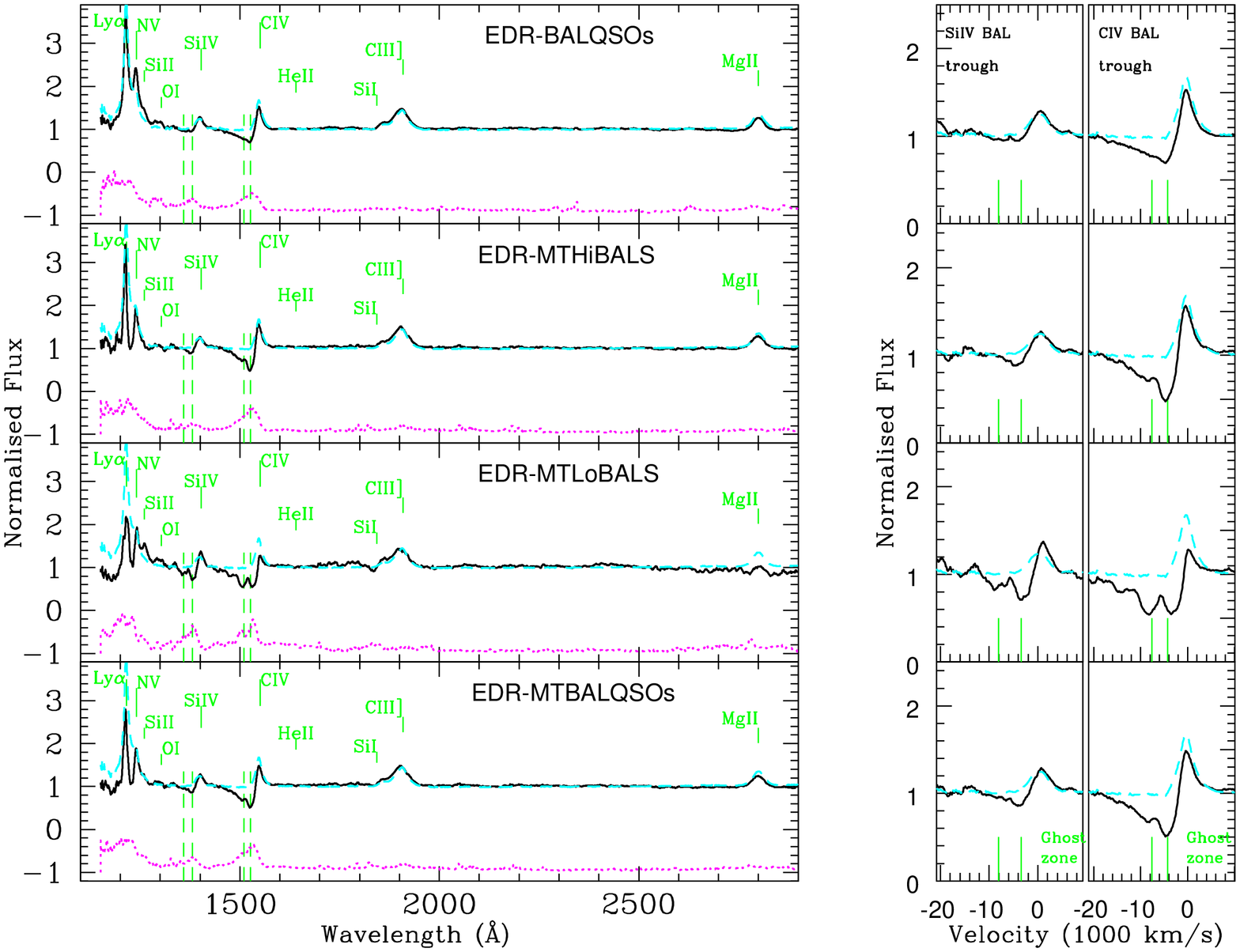}
\caption[Figure 4]{
Composite Spectra from the EDR. Black lines
display the composite spectra and blue dashed lines display the EDR
sample composite. The red dotted lines display each composites
respective RMS spectrum. As before, the vertical dashed lines trace
out the GZs in both the C$\;${\sc iv} \& Si$\;${\sc iv} BALs.
}
\label{Fig04}
\end{figure*}

\begin{figure*}
\includegraphics[scale = 0.65,angle=270]{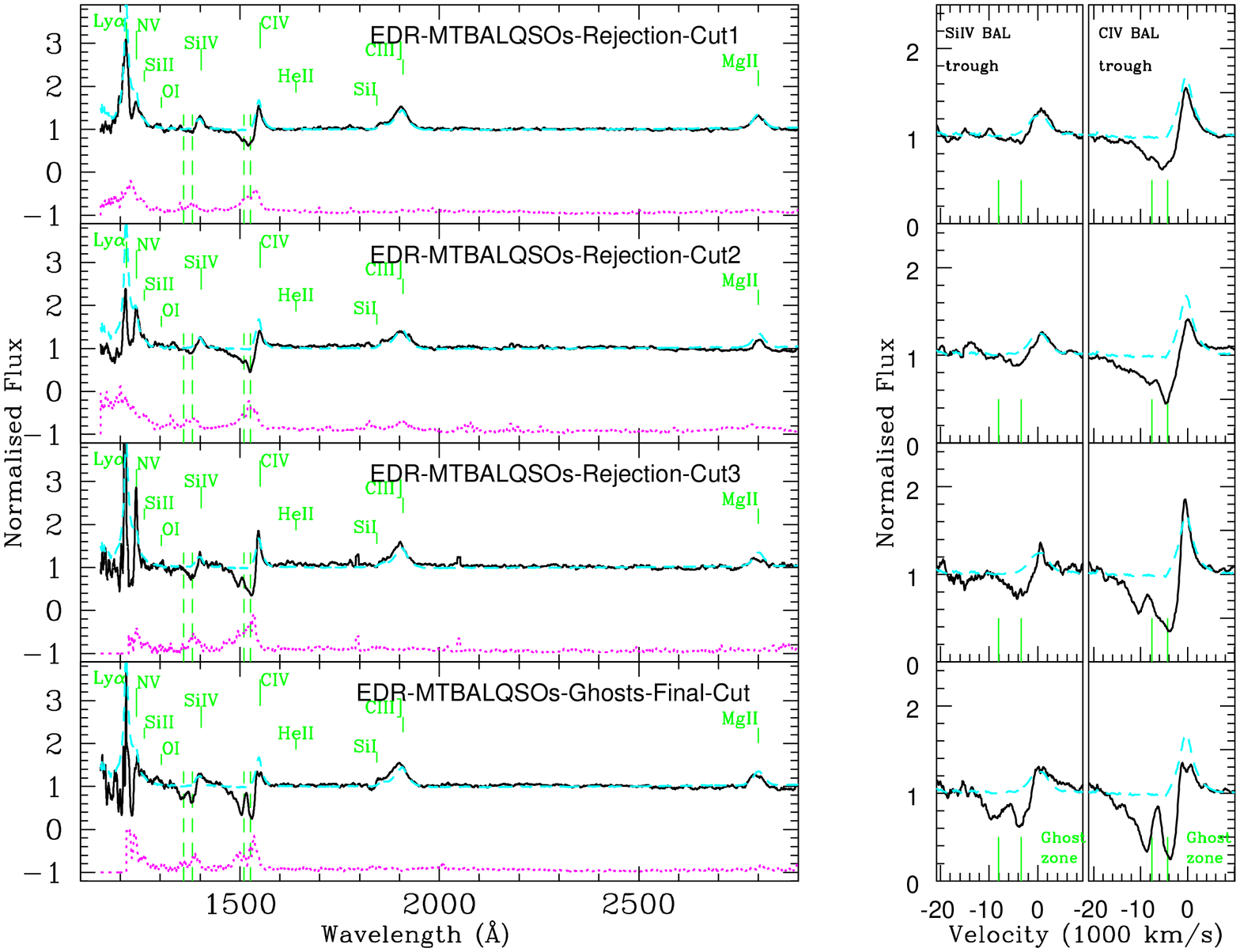}
\caption[Figure 5]{
Composite spectra from the EDR after imposing our rejection cuts. The black lines display the
composite spectra. The blue dashed lines display the EDR sample
composite. The red dotted lines display each composites respective RMS
spectrum. And, as before, the vertical dashed lines trace out the GZs
in Both the C$\;${\sc iv} \& Si$\;${\sc iv} BALs.
}
\label{Fig05}
\end{figure*}

Each composite is constructed as the arithmetic mean of the normalized
BAL QSO spectra within a given sample. The normalization is done by
fitting a composite spectrum, allowing for differences in reddening,
systematic offset and power law index, to selected continuum windows
for each source in the sample and dividing the spectrum by the
fit. This method is based upon that described by Reichard et~al.\
(2003). The resulting normalized composites are shown in
Figs.~\ref{Fig04} and~\ref{Fig05}. For comparison, we have also
constructed a composite from the full EDR QSO sample in the same way,
and this is shown overlaid on each of our BAL QSO composites.

We begin by displaying the composite of the full set of 198 BAL QSOs
satisfying our redshift constraints (Fig. 4, EDR-BAL QSOs). This
spectrum is well matched by the EDR-QSO, except within the BAL regions
themselves. This is as expected, since the EDR-QSO sample is dominated
by non-BAL QSOs with only a small admixture ($\sim$15\%) of BAL QSOs.

Fig.~\ref{Fig04}, EDR-MTHiBALs shows the composite for the EDR multi-trough
HiBALs sample. Again we see very good agreement with the EDRQSO
composite throughout the spectrum with the exception of the
BALs. However, unlike the EDR-BAL QSO full sample we do see a feature
starting to appear within the GZ of C$\;${\sc iv}. It is also
interesting to note that the C$\;${\sc iv} BAL trough is deeper in
this composite than in the full BAL QSO one. In line with this, the
trough between Ly $\alpha$ and N$\;${\sc v}, which is most likely due
to the N$\;${\sc v} BAL, is much deeper in the MTHiBAL composite than
in the full BAL QSO one. Thus, whatever the origin of the
multiple-trough structure, it appears to be associated with (or more
easily seen in) objects displaying particularly strong BALs. The
relative strength of the N$\;${\sc v} BAL trough in a composite
displaying a feature in the GZ is, of course, in line with the
dynamical model for the ghost of Ly $\alpha$. After all, in the
context of this model, the feature is due to locally enhanced
acceleration due to the scattering of Ly $\alpha$ photons by N$\;${\sc
v} ions in the flow.

Fig.~\ref{Fig04}, EDR-MTLoBALs shows the multi-trough LoBAL composite. This
spectrum contains strong C$\;${\sc iv} and Si$\;${\sc iv} BALs, with
clear features in both ghost zones. This confirms the detection of an
apparent ghost feature in the LoBAL composite presented by Reichard
et~al.\ (2003b). Closer inspection of the feature within the
Si$\;${\sc iv} GZ reveals that it is double-peaked, with a
peak-to-peak separation consistent with the Si$\;${\sc iv} doublet
separation ($\sim$2000~\kms). This is consistent with the idea that
this feature is caused by an optical depth reduction due to Ly
$\alpha$-N$\;${\sc v} line-locking.

The final composite in Fig.~\ref{Fig04}, EDR-MTBAL QSOs,
corresponds to the merged MT HiBALs and MT LoBALs samples. This
composite looks very similar to the MTHiBALs one, which is
unsurprising, since the MT HiBALs out-number the MTLoBALs by about
4:1.

We now turn to the composites for the samples generated by our various
rejection cuts, including that for our final, best-bet ghost candidate
sample. These composites provide a useful test of our selection
criteria and are shown in Fig.~\ref{Fig05}.

Fig.~\ref{Fig05}, `EDR-MTBAL QSOs-Rejection-Cut1', shows the composite
for the sample of objects rejected in Rejection Cut 1. Since none of
the objects removed by this cut individually exhibited a convincing
ghost feature, we also do not expect to see a clear feature in the GZs
of the composite. This expectation is largely confirmed although even
this composite shows a slight hint of a feature in the C$\;${\sc iv}
GZ.

The next two composites (`EDR-MTBAL QSOs-Rejection-Cut2'
and `EDR-MTBAL QSOs-Rejection-Cut3') correspond to the samples of
objects removed by Rejection Cuts 2 and 3. As already discussed above,
these cuts (especially RC3) were quite stringent, and the resulting
samples are quite likely to contain objects with genuine (and possibly
even quite strong) ghost features. It is therefore not surprising that
both samples show clearer features in the C$\;${\sc iv} GZ, and that
the RC3 composite, in particular, displays quite a strong local
maximum just beyond the blue edge of the GZ.  The off-centre location
of these features in the composites deserves additional comment. At
first sight, this would seem to be inconsistent with the ghost of Ly
$\alpha$ mechanism. However, it is important to remember that we are
dealing with composites constructed from BAL QSOs whose redshifts are
uncertain to about $\Delta{z} \sim 0.01$, whose underlying BAL troughs can be
quite asymmetric (being deeper redwards of the GZ), and whose ionisation fractions can differ
(ref. section 3, GZ definition). The combination of
slightly shifted spectra with asymmetric BALs of this type will
produce a composite in which any ghost feature is offset to the blue,
even if the ghosts in individual spectra are intrinsically at the
correct location. In the specific case of the RC3 sample, inspection
of the individual spectra in Fig.~\ref{Fig03} shows that three objects in this
sample exhibit features near the centre of the GZ, while two show
features near or just beyond the blue edge.

The final composite displayed in Fig.~\ref{Fig05} is that of our best
sample of ghost candidates (EDR-MTBAL QSOs-Ghosts-Final-Cut). This
composite shows a very clear ghost signature in C$\;${\sc iv}. Given
the way in which this sample has been selected, this is no great
surprise. However, it is encouraging that, in this composite, there is
even a feature in the GZ of Si$\;${\sc iv}.

\section{Notes on individual objects}
\label{Sec06}

In this section we present and briefly discuss the seven objects that make
up our final sample. These objects have survived all of our
rejection cuts, and we thus consider all of them to be extremely
strong ghost candidates. Table~\ref{Table01} gives additional information for each
object.

One of our goals here is to check to what extent the objects in our
sample satisfy the criteria laid out by Arav (1996) for ghost
formation. Briefly, these criteria are
\begin{enumerate}
\item The presence of a significant C$\;${\sc iv} BAL trough between $-$3000~\kms\ and $-$9000~\kms. 
\item The presence of a strong and relatively narrow Ly-$\alpha$ emission line. 
\item A clear BAL associated with N$\;${\sc v}.  
\item The power ($\nu{F_\nu}$) emitted in the region 200--1000{\AA}
should be weak compared that emitted longward of Ly$\alpha$. 
As explained below, this criterion can be tested indirectly by requiring weak/absent He$\;${\sc ii} 1640{\AA}.
\end{enumerate}

The last condition arises from the requirement that N$\;${\sc v}
should contribute a significant fraction of the total radiative
acceleration of the BAL flow. The Ly-$\alpha$ forest region contains a
result of the large number of resonant transitions, so if a lot of
energy is available in this region of the spectrum, these transitions
will contribute strongly to the total driving force. As noted by Arav
(1996), the equivalent width of He$\;${\sc ii} 1640{\AA} is known to
correlated with the intrinsic flux at the He$\;${\sc ii} ionization
energy at 228{\AA}, so weak or absent He$\;${\sc ii} 1640{\AA} can
serve as a convenient proxy for the last criterion.

\begin{figure*}
\includegraphics[scale = 0.65,angle=270]{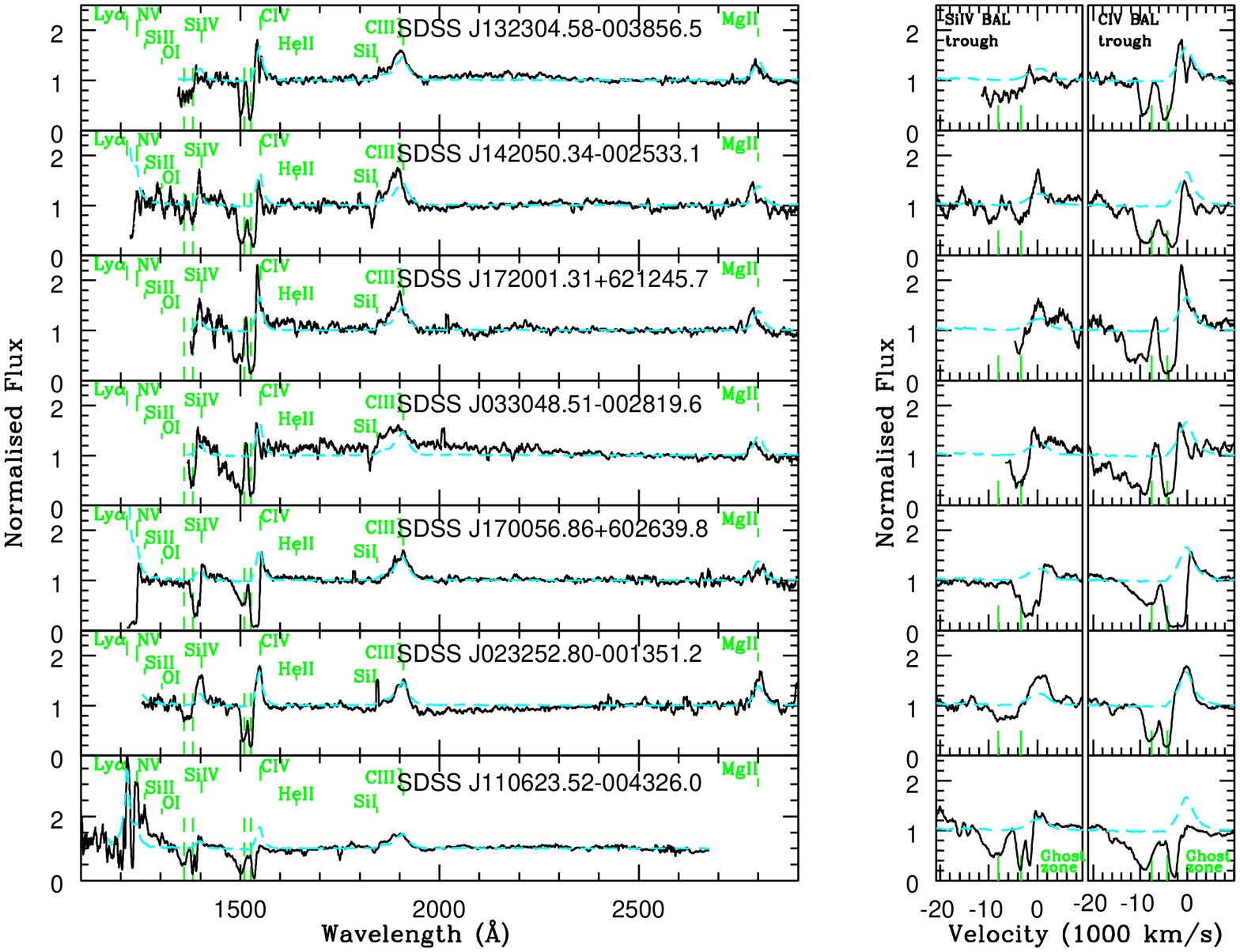}
\caption[Figure 6]{
Individual objects from our 'Ghost Candidates Final Cut' set. Here, the black lines display the
composite spectra. The blue dashed lines display the EDR sample
composite. And, as before, the vertical dashed lines trace out the GZs
in both the C$\;${\sc iv} and Si$\;${\sc iv} BALs.
}
\label{Fig06}
\end{figure*}

Having laid out the criteria for the formation of the ghost of Ly
$\alpha$, we now  briefly discuss each of our
strong ghost candidates, displayed in Fig.~\ref{Fig06}.

\setcounter{subsection}{1}
\subsubsection{SDSS J132304.58$-$003856.5} (HiBAL)
We first note the lack of Ly $\alpha$ and N$\;${\sc v} coverage for
this object, which prevents us from testing criteria 2 and 3. Criteria
1 and 4 are clearly met, however. The ghost feature itself is very
pronounced and surrounded by a strong, deep BAL trough on both sides.
The local maximum is located closer to the blue edge of the GZ, but
inspection of the C$\;${\sc iii}] and Mg$\;${\sc ii} BELs suggests
that the redshift may be at fault. If the redshift is adjusted so that
these two BELs are found at their expected locations, the ghost
feature is also better centred in the GZ.

\subsubsection{SDSS J142050.34$-$002553.1} (LoBAL)
This object again shows a strong C$\;${\sc iv} BAL and a clear feature centred
in the ghost zone. However, Mg$\;${\sc ii} and C$\;${\sc iii}] again
appear to be somewhat misaligned, and in this instance centring on
these lines shifts the peak of the ghost feature further to the red
edge of the GZ. The Si$\;${\sc iv} BAL shows a deep trough either side of the
Ghost feature. Shifting the spectrum to centre Mg$\;${\sc ii} and
C$\;${\sc iii}] serves to centre this feature in the middle of the
GZ. Closer inspection of the Si$\;${\sc iv} feature also reveals the expected
double peak structure.  Again there appears to be little or no trace
of a He$\;${\sc ii} 1640{\AA}. There is no coverage of Ly $\alpha$, but
the steep drop near the blue end of the spectrum probably corresponds
to the red end of the N$\;${\sc v} BAL. Thus we cannot test for
criterion 2, but criteria 1, 3 and 4 are satisfied by this object.

\subsubsection{SDSS J172001.31+621245.7} (HiBAL)
Again there is a lack of coverage of both Ly $\alpha$ and N$\;${\sc v}
for this object. The C$\;${\sc iv} BAL is strong and broad, and exhibits a clear
local maximum in the GZ. This feature again sits close to the blue
edge of the GZ, but a redshift adjustment that centres Mg$\;${\sc ii}
again centres the ghost feature also. There is no obvious He$\;${\sc
ii} emission. Thus two of the four criteria are clearly met, while the
other two cannot be tested.

\subsubsection{SDSS J033048.51$-$002819.6} (HiBAL)
All of the comments just made for SDSS J172001.31+621245.7 also apply
here. This includes the improvement in the position of the ghost
feature when the redshift is optimized to centre the Mg$\;${\sc ii}
line. Again, two criteria are met, and two cannot be tested.

\subsubsection{SDSS J170056.85+602639.8} (HiBAL)
This object displays a spectrum that cuts off near N$\;${\sc v}. There
is no coverage of Ly $\alpha$, but, as in the case of SDSS
142050.34-002553.1, the steep drop at the blue end of the spectrum
suggests the presence of a strong N$\;${\sc v} BAL. The candidate
ghost feature is well centred in the C$\;${\sc iv} GZ, in line with the fact
that Mg$\;${\sc ii} and C$\;${\sc iii}] appear to be well
centred. There is no obvious He$\;${\sc ii} 1640{\AA} emission, so
criteria 1, 3 and 4 are met and only criterion 2 (presence of strong
Ly $\alpha$) cannot be tested.

\subsubsection{SDSS J023252.80$-$001351.2} (HiBAL)
The candidate ghost feature in the object is located in a deep, clear
C$\;${\sc iv} BAL trough and is well centred in the GZ. Mg$\;${\sc ii} and
C$\;${\sc iii}] are also well centred, confirming Reichard et al.'s
(2003) redshift estimate. There is again no coverage of Ly $\alpha$
and N$\;${\sc v}, but also no sign of He II 1640{\AA}. Thus two of the
ghost criteria are met, and two cannot be tested.

\subsubsection{SDSS J110623.52$-$004326.0} (LoBAL)
Our final ghost candidate is the only object in this sample that does
have coverage of the Ly $\alpha$ BEL. The line is certainly strong,
although its breadth is difficult to judge due to the presence of
other transitions (including N$\;${\sc v}). A strong N$\;${\sc v} BEL
is also present and accompanied by a deep BAL. The C$\;${\sc iv} BAL is strong
and deep, and contains a clear local maximum that is centred in the
GZ. The spectral coverage does not extend to Mg$\;${\sc ii}, but
C$\;${\sc iii}] appears to be reasonably well centred. This object
also shows a clear feature in the Si$\;${\sc iv} GZ, but there is also another,
narrower local maximum redwards of the putative ghost feature in this
line. All of the ghost criteria are satisfied making this our
strongest candidate.

\begin{table*}
\caption[]{Quasars in the Multiple-Trough sample.  Categories (final column) are
Ghost Candidate (GC), Possible Ghost (PG), and Final Cut (FC),
with RC1--3 indicating the `rejection cut' samples of
Section~\ref{SecMeth}.}

\begin{tabular}{lrlllc}
\hline
\multicolumn{1}{c}{Object ID} &
Reichard's BI &
Redshift &
Magnitude &
S/N &
Category\\
&
&
\multicolumn{1}{c}{$z$}&
$(g)$&
&
\\
\hline

J110623.52$-$004326.0 &
 4034 &              
 2.450 &             
 19.38 &             
 10.140 &            
 GC : FC\\           
                     
J023252.80$-$001351.2 &
 2092 &              
 2.025 &             
 19.28 &             
 10.160 &            
 GC : FC\\           
                     
J170056.85+602639.8 &
 1400 &              
 2.125 &             
 19.24 &             
 9.427 &             
 GC : FC\\           
                     
J033048.51$-$002819.6 &
 5548 &              
 1.779 &             
 19.62 &             
 6.593 &             
 GC : FC\\           
                     
J172001.31+621245.7 &
 3290 &              
 1.762 &             
 19.47 &             
 5.781 &             
 GC : FC\\           
                     
J142050.34$-$002553.1 &
 3442 &              
 2.103 &             
 19.90 &             
 4.345 &             
 GC : FC\\           
                     
J132304.58$-$003856.5 &
 287 &               
 1.828 &             
 18.64 &             
 8.491 &             
 GC : FC\\           
                     
J143022.47$-$002045.2 &
 1957 &              
 2.544 &             
 20.72 &             
 2.473 &             
 PG : RC3\\          
                     
J145045.42$-$004400.3 &
 238 &               
 2.078 &             
 18.59 &             
 14.120 &            
 PG : RC3\\          
                     
J171330.98+610707.8 &
 0 &                 
 1.685 &             
 19.10 &             
 8.611 &             
 PG : RC3\\          
                     
J113544.33+001118.6 &
 3379 &              
 1.723 &             
 20.32 &             
 4.958 &             
 PG : RC3\\          
                     
J110736.67+000329.4 &
 123 &               
 1.740 &             
 18.65 &             
 18.570 &            
 PG : RC3\\          
                     
J005355.15$-$000309.3 &
 1088 &              
 1.715 &             
 18.60 &             
 13.230 &            
 PG : RC2\\          
                     
J010616.05+001523.9 &
 2520 &              
 3.050 &             
 20.46 &             
 2.649 &             
 PG : RC2\\          
                     
J010612.21+001920.1 &
 2453 &              
 3.110 &             
 19.19 &             
 7.406 &             
 PG : RC2\\          
                     
J020006.31$-$003709.7 &
 9550 &              
 2.136 &             
 18.81 &             
 14.020 &            
 PG : RC2\\          
                     
J025042.45+003536.7 &
 3544 &              
 2.380 &             
 19.29 &             
 7.418 &             
 PG : RC2\\          
                     
J100809.63$-$000209.9 &
 56 &                
 2.561 &             
 19.24 &             
 5.162 &             
 PG : RC2\\          
                     
J104109.85+001051.8 &
 1913 &              
 2.250 &             
 19.14 &             
 12.090 &            
 PG : RC2\\          
                     
J104233.86+010206.3 &
 401 &               
 2.123 &             
 18.93 &             
 14.020 &            
 PG : RC2\\          
                     
J104841.02+000042.8 &
 1176 &              
 2.022 &             
 18.91 &             
 12.730 &            
 PG : RC2\\          
                     
J120657.01$-$002537.8 &
 110 &               
 2.005 &             
 19.45 &             
 6.306 &             
 PG : RC2\\          
                     
J123947.61+002516.2 &
 7299 &              
 1.869 &             
 20.27 &             
 2.972 &             
 PG : RC2\\          
                     
J130035.29$-$003928.4 &
 853 &               
 3.630 &             
 20.28 &             
 2.994 &             
 PG : RC2\\          
                     
J134544.55+002810.8 &
 1510 &              
 2.516 &             
 18.83 &             
 11.070 &            
 PG : RC2\\          
                     
J134808.79+003723.2 &
 1309 &              
 3.620 &             
 20.36 &             
 2.697 &             
 PG : RC2\\          
                     
J143054.03$-$003627.3 &
 9064 &              
 3.710 &             
 22.31 &             
 0.372 &             
 PG : RC2\\          
                     
J145913.72+000215.8 &
 356 &               
 1.910 &             
 18.63 &             
 12.990 &            
 PG : RC2\\          
                     
J151636.79+002940.4 &
 4035 &              
 2.240 &             
 18.48 &             
 12.650 &            
 PG : RC2\\          
                     
J171944.76+554408.3 &
 205 &               
 3.886 &             
 21.74 &             
 1.230 &             
 PG : RC2\\          
                     
J171949.92+532132.8 &
 4903 &              
 1.777 &             
 18.22 &             
 16.940 &            
 PG : RC2\\          
                     
J173911.52+565550.9 &
 919 &               
 1.772 &             
 19.28 &             
 8.888 &             
 PG : RC2\\          
                     
J234506.32+010135.5 &
 2488 &              
 1.794 &             
 19.70 &             
 7.732 &             
 PG : RC2\\          
                     
J134145.13$-$003631.0 &
 870 &               
 2.205 &             
 19.57 &             
 8.652 &             
 PG : RC2\\          
                     
J000056.89$-$010409.8 &
 1560 &              
 2.111 &             
 20.41 &             
 3.437 &             
 PG : RC2\\          
                     
J143022.47$-$002045.2 &
 1957 &              
 2.544 &             
 20.72 &             
 2.473 &             
 PG : RC2\\          
                     
J003551.98+005726.3 &
 1731 &              
 1.905 &             
 19.24 &             
 8.533 &             
 PG : RC1\\          
                     
J004041.39$-$005537.3 &
 0 &                 
 2.092 &             
 18.18 &             
 17.240 &            
 PG : RC1\\          
                     
J011227.60$-$011221.7 &
 3033 &              
 1.755 &             
 18.12 &             
 16.320 &            
 PG : RC1\\          
                     
J012913.70+011428.0 &
 345 &               
 1.782 &             
 19.50 &             
 6.105 &             
 PG : RC1\\          
                     
J015048.82+004126.2 &
 105 &               
 3.703 &             
 19.76 &             
 5.479 &             
 PG : RC1\\          
                     
J024221.86+004912.7 &
 229 &               
 2.071 &             
 18.54 &             
 16.480 &            
 PG : RC1\\          
                     
J031227.13$-$003446.2 &
 0 &                 
 1.772 &             
 19.72 &             
 7.315 &             
 PG : RC1\\          
                     
J104152.61$-$001102.1 &
 1588 &              
 1.703 &             
 19.21 &             
 12.580 &            
 PG : RC1\\          
                     
J110041.19+003631.9 &
 4687 &              
 2.017 &             
 18.62 &             
 15.030 &            
 PG : RC1\\          
                     
J121803.28+001236.8 &
 269 &               
 2.010 &             
 19.44 &             
 8.181 &             
 PG : RC1\\          
                     
J122228.39$-$011011.0 &
 678 &               
 2.284 &             
 19.74 &             
 5.290 &             
 PG : RC1\\          
                     
J123124.71+004719.1 &
 3134 &              
 1.720 &             
 19.57 &             
 5.513 &             
 PG : RC1\\          
                     
J123824.90+001834.5 &
 220 &               
 2.154 &             
 19.30 &             
 6.652 &             
 PG : RC1\\          
                     
J130348.94+002010.4 &
 1425 &              
 3.655 &             
 20.77 &             
 1.525 &             
 PG : RC1\\          
                     
J170903.06+594530.7 &
 4936 &              
 1.708 &             
 19.18 &             
 10.250 &            
 PG : RC1\\          
                     
J170931.00+630357.1 &
 0 &                 
 2.402 &             
 18.41 &             
 13.090 &            
 PG : RC1\\          
                     
J170951.03+570313.7 &
 528 &               
 2.547 &             
 20.95 &             
 2.505 &             
 PG : RC1\\          
                     
J172012.40+545601.0 &
 1249 &              
 2.099 &             
 18.47 &             
 21.920 &            
 PG : RC1\\          
                     
J232205.46+004550.9 &
 222 &               
 1.820 &             
 20.55 &             
 3.032 &             
 PG : RC1\\          
                     
J032246.82$-$005148.9 &
 0 &                 
 1.680 &             
 19.60 &             
 7.193 &             
 PG : RC1\\          
                     
J130208.26$-$003731.6 &
 0 &                 
 1.672 &             
 18.44 &             
 13.510 &            
 PG : RC1\\          
                     
J125241.55$-$002040.6 &
 2524 &              
 2.898 &             
 18.90 &             
 10.230 &            
 PG : RC1\\          

\hline
\label{Table01}
\end{tabular}
\end{table*}

\begin{figure*}
\includegraphics[scale = 0.65,angle=270]{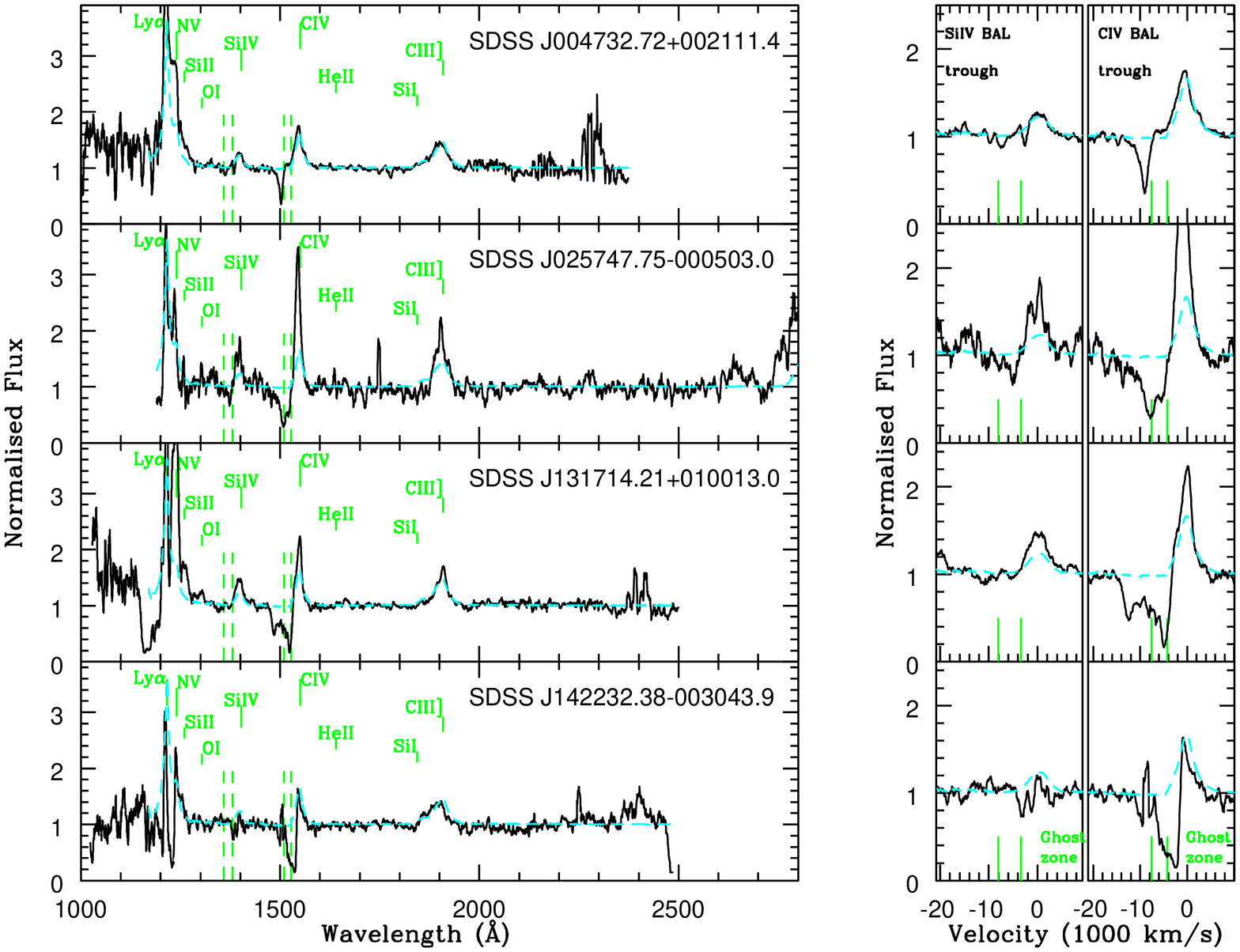}
\caption[Figure 7]{
Examples of EDR BAL QSOs, not present in our GCFC sample, that were tested for Arav's criteria.
Here, the black lines display the composite spectra. The blue dashed lines display the EDR sample
composite. And, as before, the vertical dashed lines trace out the GZs
in both the C$\;${\sc iv} and Si$\;${\sc iv} BALs.
}
\label{Fig07}
\end{figure*}

\section{Discussion}

Of the 198 SDSS EDR BAL QSOs identified by Reichard et al.\ (2003),
33  are identified as LoBALs and 165 as
HiBALS. This is consistent with the incidence of HiBALs and LoBALs
amongst the BAL QSO population (Sprayberry \& Foltz 1992).  We find
that $F_{\rm g}$, the fraction of BAL QSOs displaying clear ghost signatures,
lies in the range $7/198 \le F_{\rm g} \le 36/198$. This corresponds to a ghost
frequency of $0.15 F_{\rm g}$ (a few per cent) amongst QSOs in
general. While it is of
interest to compare the relative fractions of ghosts amongst HiBALs
and LoBALs (5/165 and 2/33 respectively), the small-number statistics
preclude us from drawing any firm conclusions at this time concerning their
likely incidence.\footnote{The classification of J142050.34$-$002553.01 as a
LoBAL is uncertain (Reichard et al. 2003)}
 This question will be addressed in a forthcoming
analysis of BAL QSOs in the SDSS DR3/4 releases.  Whilst we have
endeavored to test Arav's selection criteria for ghost candidates (see
Section~\ref{Sec06}), this was not possible in every circumstance. We note that
adherence to Arav's stringent selection criteria will only be possible,
whilst providing good statistics, with the far larger SDSS DR3/4
datasets. However, we note that none of our Ghost Candidate Final Cut
(GCFC) set violates Arav's selection criteria (in so far as they can
be verified) for ghost candidates.

As a final sanity check, we felt it important to perform two additional
tests. The first was to turn Arav's theorem
around and ask if there are objects in the SDSS EDR sample that
satisfy all of the criteria but do not exhibit ghost features. If the
line-locking interpretation for the origin of the ghost is correct,
such objects should not exist (Arav 1996). We have carried out this
exercise and inspected by eye all 74 objects in the SDSS-EDR BAL QSO
sample, but not present in our GCFC set, that contain the region
between Lyman alpha and He$\;${\sc ii} that is the minimum necessary
to test all the criteria. We find that 56 of these 74 BAL QSOs clearly
do not satisfy at least one of Arav's selection criteria
(e.g. Fig. 7, SDSS J004732.72+002111.4). The remaining 18  appear to
be borderline cases, i.e., they exhibit a degree of ambiguity in just 
one of the criteria. All of these objects were cut at RC1. Of these,
most show the C$\;${\sc iv} BAL returning to continuum level at around
$-$9000~\kms\ but do contain the slightest hint
of a ghost (e.g. Fig. 7, SDSS J025747.75$-$000503.0). In line with this
is the appearance of a relatively weak
N$\;${\sc v} BAL; thus only a weak ghost should, perhaps, be expected.
Of the remaining MT BAL QSOs, a couple actually appear to satisfy all
criteria but show no ghost (e.g. Fig. 7, SDSS J131714.21+010013.0).
However, we note that for these two
objects there is relatively weak continuum emission compared to the
red end of the Ly $\alpha$ forest and thus Arav's criteria still
hold. This does, however, cast some doubt on the reliability of using
He$\;${\sc ii} emission as a proxy for Ly-$\alpha$ forest flux. One
final object in this set of 18 BAL QSOs is the clearest object that
`should' reveal a Ghost (ref. Fig. 7, SDSS J142232.38$-$003043.9).
However, this was rejected at RC1 purely
on the basis that it possesses an additional emission feature just
blueward of the ghost zone (most likely a residual sky line), which
obscures what otherwise appears to be a genuine ghost.  Evidently,
these results show a small degree of ambiguity, but again, the vast
numbers of DR3/DR4 will enable us to provide robust evidence as to the
validity of this hypothesis. 

The second test we performed was to confirm that
there really is an excess of objects with "bumps" in the ghost zone. 
To achieve this we have carried out the following. The GZ was 
systematically offset by 2000~\kms\ both to the red and to the blue 
of its correct location. The full analysis of the MTBAL QSO set was 
then reperformed for these two new locations. With the GZ shifted 
artificially to the red there were no objects that made it to the GCFC 
stage. Clearly, the distribution of objects within each rejection cut 
stage was altered but the net effect was to find no `good' ghost candidates. 
With the GZ shifted artificially to the blue a GCFC set of six objects 
was produced. Of these 6 objects, 4 are from our original RC3 sample, 1 is 
from the original GCFC sample and only 1 new object was introduced. Again, 
as for the red-shifted GZ the distributions amongst the rejection cut stages 
were altered slightly. However, this analysis clearly confirms that the 
"bumps" in the absorption troughs of our best ghost candidates are likely to 
be genuine ghosts of Lyman alpha.

\section{Conclusions}

We have searched Reichard et al.'s (2003) BAL QSO catalogue, based on
the SDSS EDR, for objects displaying clear `ghost of Ly $\alpha$'
signatures. To this end, we have carried out several stages of
rejection, constructing a number of sub-samples along the way. Since
our selection criteria are quite strict, some of the features we have
rejected along the way may nevertheless be genuine ghosts (and this is
especially true for the 5 objects rejected in our third and final cut,
i.e., the RC3 sample).

Our very best sample contains seven objects that have survived all of our
cuts. All of these display strong and broad C$\;${\sc iv} BALs, and all exhibit
clear local maxima at the locations expected for the ghost
signature. None of them have been found to violate any of the criteria
laid out by Arav (1996) for the formation of the ghost of Ly
$\alpha$. However, in most cases the limited wavelength coverage of
the data prevents us from testing for all criteria
simultaneously. Nevertheless, we believe that all objects in this
sample are excellent ghost candidates.

It is our hope that this paper will encourage follow-up observations
and detailed modelling of the objects in our sample. After all, the
ghost of Ly $\alpha$ represents the clearest observational signature
of (and the only direct evidence for) the mechanism that powers
outflows from (BAL) QSOs. As such, it has been underexploited. For
example, the variability properties of the ghost signature remain
completely unknown at present. We plan to rectify this in the near
future, using our new sample of objects as a basis. The goal of this
program will be to test if ghost variability can be used to gain
insight into the BAL region, in the same way that classical
reverberation mapping has yielded key information regarding the nature
of the BEL region.

\section*{Acknowledgements}
We thank Professor Kirk Korista for his valuable input into the production of this paper
and also thank Dr Gordon Richards for his positive and helpful contribution.

\end{document}